\pdfoutput=1
\documentclass[twocolumn]{aastex631}

\let\tablenum\relax
\usepackage{siunitx}

\usepackage{amsmath,amssymb}
\usepackage{graphicx}
\usepackage{longtable}
\usepackage{tabularx}
\usepackage{verbatim}
\usepackage{tikz}
\tikzset{>=latex}

\setcitestyle{notesep={ }}

\usepackage{cleveref}
\crefname{figure}{Figure}{Figures}
\crefname{equation}{Equation}{Equations}
\crefname{table}{Table}{Tables}
\crefname{section}{Section}{Sections}
\crefname{subsection}{Subsection}{Subsections}
\crefformat{equation}{Equation~#2#1#3}
\crefformat{section}{Section~#2#1#3}
\crefformat{subsection}{Subsection~#2#1#3}

\hypersetup{
  colorlinks=true,
  linkcolor=blue,
  filecolor=magenta,
  urlcolor=cyan,
  citecolor=blue
}

\shorttitle{Bouhrik et al.}
\shortauthors{Bouhrik et al.}

\begin{document}

\title{DISCOVERY AND MULTI-WAVELENGTH ANALYSIS OF A NEW DISSOCIATIVE GALAXY CLUSTER MERGER: THE CHAMPAGNE CLUSTER}

\author{Faik Bouhrik}

\affiliation{University of California Davis \\
One Shields Ave \\ Davis, CA 95616, USA}

\author{Rodrigo Stancioli}
\affiliation{University of California Davis \\
One Shields Ave \\ Davis, CA 95616, USA}

\author{David Wittman}
\affiliation{University of California Davis \\
One Shields Ave \\ Davis, CA 95616, USA}

\begin{abstract}

We report the discovery of a new binary galaxy cluster merger, the Champagne Cluster (RM J130558.9+263048.4), using a detection method that identifies dynamically active clusters in the redMaPPer SDSS DR8 photometric galaxy cluster catalog. The Champagne Cluster exhibits the classic X-ray morphology of a post-pericenter dissociative galaxy cluster merger: an X-ray peak located between two galaxy overdensities at the same redshift. We conducted a Keck/DEIMOS survey and obtained redshifts for {\bfseries 102} member galaxies. The redshift analysis indicates a relative velocity of 411 $\pm$ 180 km/s between the two subclusters, which suggests that the merger is happening near the plane of the sky. We estimated the bulk temperature (8.20 $\pm 1.2$ keV) and total X-ray luminosity (7.29 $\pm$ 0.19 $\times$ $10^{44}$ erg  $\times$ $s^{-1}$) of the intracluster medium using $\textit{Chandra}$ archival data. We used the $\textit{ClusterPyXT}$ software to make a temperature map, and we compared it to hydrodynamic simulations to constrain the time since pericenter (TSP), the impact parameter, and the mass ratio. We found two scenarios that matched our data: a returning system with an impact parameter of 0 and TSP of 2.2 Gyr, and an outbound system with an impact parameter of 500 kpc and TSP of 0.4 Gyr. Both scenarios have a mass ratio of 1:10.

\end{abstract}


\section{Introduction} \label{sec:intro}

According to the $\Lambda$CDM model of cosmology, density fluctuations in the primordial universe collapsed to form the seeds of cosmic structure. These seeds hierarchically grew to form larger and larger structures through merging and accretion, eventually building galaxy clusters \citep{2012AnP...524..507F}. In recent decades, galaxy clusters have been established as powerful probes for a host of astrophysical and particle physics processes \citep{doi:10.1146/annurev-astro-081811-125502}: the hierarchical formation of large structure in the universe \citep{COHN2005316}, rates of star formation in dense environments \citep{ 2003AJ....125.2427M,10.1093/mnrasl/slx041}, and the nature and properties of dark matter \citep{Markevitch04, bradac2008, Dawson11, Wittman_2023}.

In a major galaxy cluster merger, the three components (dark matter halos, galaxies, and the intracluster medium (ICM)), may become separated \citep{2006ApJ...648L.109C}. Offsets between dark matter, galaxies, and X-ray surface brightness peaks offer rare glimpses into the properties of dark matter, allowing for the constraining of the scattering cross-section for dark matter particles \citep{Markevitch04}. It was such an offset exhibited by the iconic Bullet Cluster (1E 0657-56) that provided the first direct evidence for the existence of dark matter and allowed for imposing an upper limit on the self-interaction cross-section for dark matter particles \citep{2006ApJ...648L.109C,Randall2008}. In the following years, only a handful of such clean bimodal systems were identified, for example A520 \citep{2007}, MACS J0025.4-1222 \citep{bradac2008}, A1758 \citep{2008PASJ...60..345O}, and DLSCL J0916+2953 \citep{Dawson11}.

We have been working on a method based on galaxy cluster bimodality, which has allowed for a more targeted search than serendipitous disturbed X-ray morphology detection. We build on the the redMaPPer algorithm \citep{Rykoff2014, Rykoff_2016}, which is a photometric red-sequence galaxy cluster finder designed specifically for large photometric surveys, such as the Sloan Digital Sky Survey (SDSS) DR8 \citep{2011ApJS..193...29A}. redMaPPer identified 26,111 galaxy cluster candidates in the SDSS DR8 catalog. We select clusters that are not dominated by a single brightest cluster galaxy (BCG), and where the minimum angular separation between the top two BCG candidates is at least $\sim$$1^\prime$. We chose this limit because it is multiple times the size of the angular resolution for XMM-$\textit{Newton}$ ($10^{\prime\prime}$), which allows us to pinpoint the location of the X-ray surface brightness between the BCGs with accuracy and because $\sim$$1^\prime$ is about double the BCG separation of the median cluster. Clusters that satisfy these two criteria become candidates for X-ray $\textit{Chandra}$ and XMM-\textit{Newton} archival searches. The location of the X-ray peak between the top two BCG candidates made some merger candidates of particular interest for immediate study, as a dissociative X-ray morphology is indicative of a post pericenter merger that has an axis component in the plane of the sky. Such systems can play a pivotal role in constraining dark matter properties. RM J130558.9+263048.4, the subject of this paper, is one of these candidates. Two other candidates that were discovered using the same method are Abell 56 \citep{Wittman_2023} and RM J150822.0+575515.2 \citep{Stancioli}. 

In this paper, we assume a $\Lambda$CDM cosmology with a $H_{o}$ = 69.6 km/s and $\Omega_{o}$ = 0.286. Unless otherwise stated, all quoted uncertainties correspond to the 68\% (1$\sigma$) confidence level.

\section{The Champagne Cluster: Initial Overview}
\label{sec:overview}

For easy reference, we assign the nickname the Champagne Cluster to RM J130558.9+263048.4. redMaPPer includes in its richness calculation the total number of probable cluster members, which leads to a richness of 70 for the Champagne Cluster \citep{Rykoff_2016}. The Champagne Cluster is poorer than most binary clusters known to the literature, however it is still richer than 93.7$\%$ of the members of the redMaPPer catalog. We estimate its mass ($M_{200}$) to be $4.65^{+11.50}_{-4.69}$ $\times$ $10^{14}$ $h^{-1}$ $M_\odot$) using the mass-richness relation for the redMaPPer catalog, where the mass was calibrated via SDSS weak lensing data \citep{10.1093/mnras/stw3250}. SDSS DR8 assigns the spectroscopic redshifts of 0.30812 and 0.30815 to the top two BCG candidates, as indicated in Table \ref{table:1}. Using the relative velocities of the two BCGs as proxies for the relative velocities of the subclusters yields a velocity of one cluster in the frame of the other, $\Delta$$v_{los}$, of only 6 $\pm$ 20 km/s. This very low line-of-sight velocity hints at a merger that is happening nearly in the plane of the sky and/or a merger that is at turnaround. In $\S$\ref{section:Manual} we refine this estimate by using the average redshift for each subcluster as a proxy for the subcluster redshift, and hence velocity. 

The angular separation between the BCGs of the two subclusters is $1.^{\prime}$17, which translates to a physical separation of 335 kpc at the cluster’s redshift.

\begin{figure*}
  \centering
  \includegraphics[width=180mm]{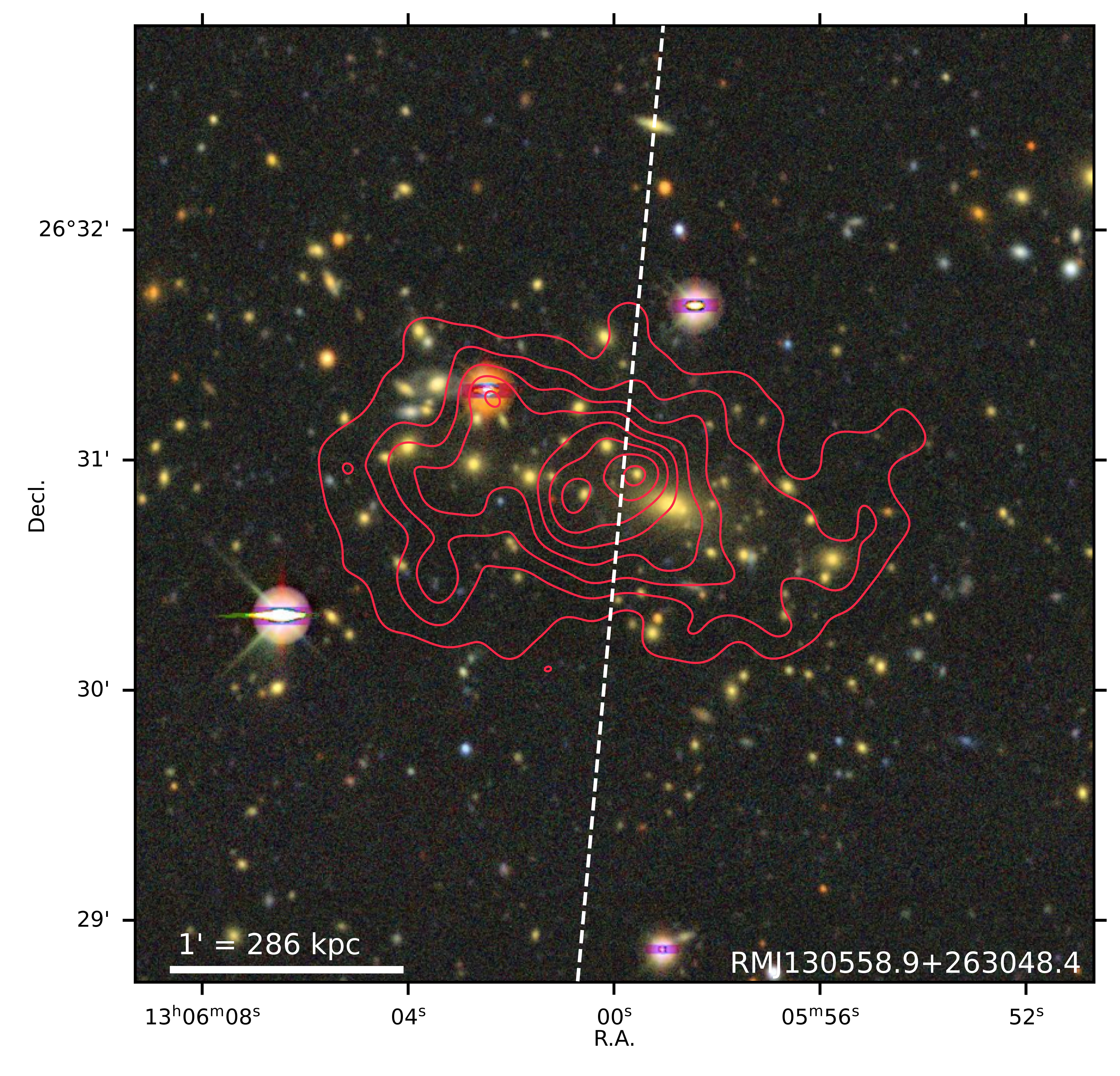}
  \caption{$\textit{Chandra}$ X-ray surface brightness contour map for the Champagne Cluster overlaid on a Legacy Survey color image. The X-ray surface brightness peak is located along the axis that connects the top two BCG candidates. The dashed line separates the two subclusters: Champagne-SE and Champagne-NW. For creating the X-ray image, the energy band 0.5-7.0 keV was selected. Point sources were removed, and the X-ray contours were derived using a Gaussian kernel with a standard deviation of $6.88^{\prime\prime}$.}
  \label{fig: Champagne}
\end{figure*}

In Figure~\ref{fig: Champagne}, we show an X-ray surface brightness contour map overlaid on a DESI Legacy Survey DR10 color image. The location of the X-ray peak along the axis that connects the top two BCG candidates is a good indicator of a post pericenter galaxy cluster merger. 

\begin{table}
\caption{Top two BCG candidates for the Champagne Cluster}              
\label{table:1}      
\centering                                      
\begin{tabular}{c c c c}          
\hline\hline                        
BCG & R.A. & Decl. & z\\    
\hline                                   
    Top BCG candidate & 196.4954 & 28.150217 & 0.30812 \\      
    2nd BCG candidate & 196.5167 & 28.143211 & 0.30815 \\

\hline                                             
\end{tabular}
\end{table}

\begin{figure}
  \centering
  \includegraphics[width=\columnwidth]{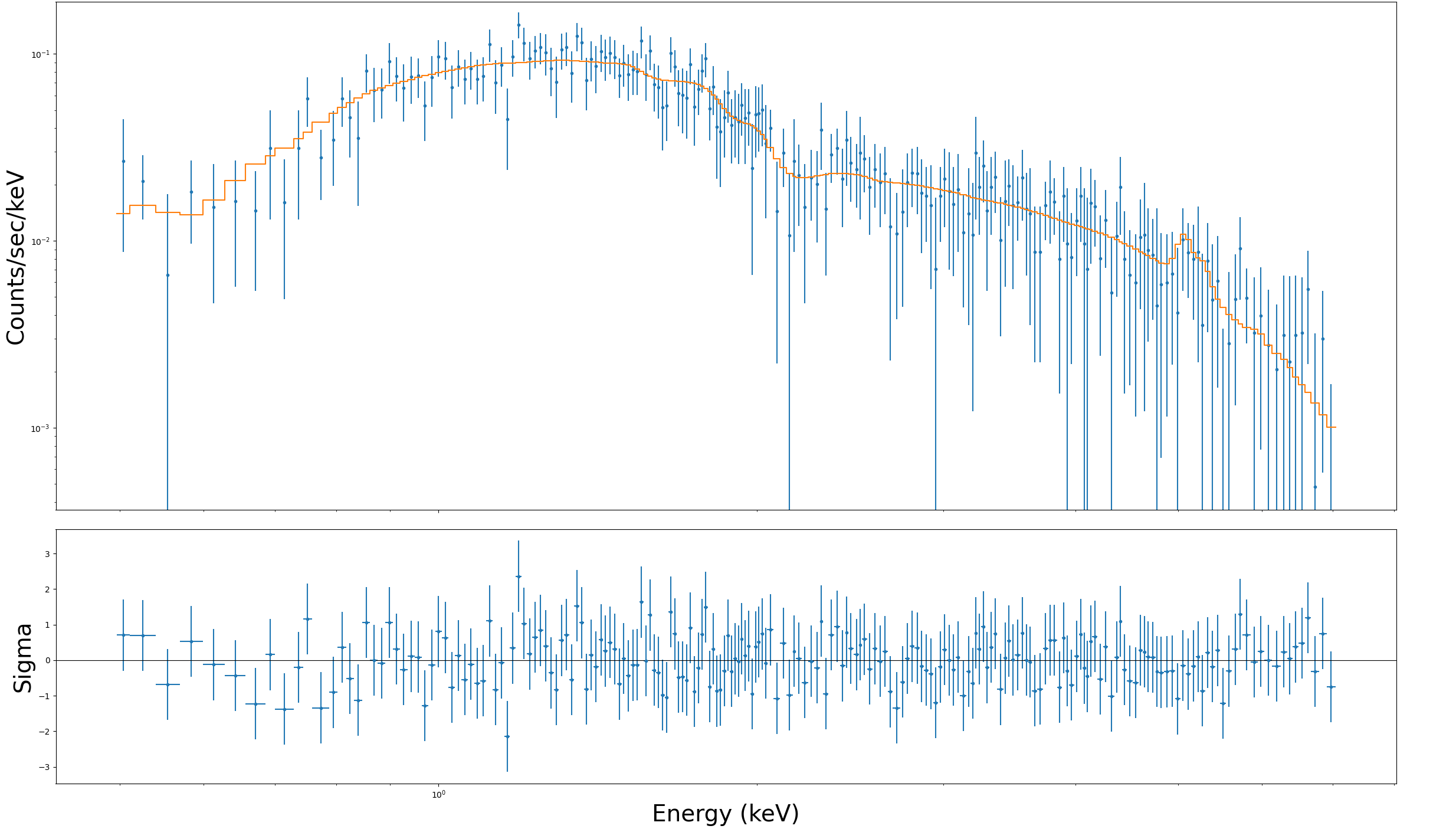}
  \caption{Top panel: The best fit model {\bfseries (apec * phabs)} for the $\textit{Chandra}$ X-ray spectrum of the Champagne Cluster. The bulk temperature corresponding to the model is 8.28 $\pm 1.1$ KeV. Bottom panel: Spectrum fitting residuals.}
  \label{fig:spectrum}
\end{figure}

The Champagne Cluster (Plank cluster name: PSZ2 G023.17+86.71) is one of the 83 Planck clusters that host extended radio emissions (radio halos) at the locations of the clusters in the LoTSS-DR2 catalog  \citep{Botteon_2022}. The halo of the Champagne Cluster extends in the West-East direction, and there are no radio relics associated with it. 

\section{$\textit{Chandra}$ data reduction and analysis}
\subsection{Spectral Fitting}
\label{sec: Chandra}
The Champagne Cluster was observed by $\textit{Chandra}$ on 2012 November 18 (Observation ID: 14015, PI: Cassano). The exposure time was 29500 s. We used CIAO version 4.14 \citep{2006SPIE.6270E..1VF} to analyze the data. We removed the point sources in the field using $\texttt{wavdetect}$, a Mexican-hat wavelet source detection tool, before estimating the global temperature ($T_X$) and luminosity ($L_X$). We chose $\texttt{wavdetect}$ as our subtraction tool because of its ability to separate closely-spaced point sources, a trait that helps in distinguishing closely separated point sources from extended sources. To extract a spectrum, we enclosed the extended emission area by an elliptical aperture with a semi-major axis of \ang{;;130} and a semi-minor axis of \ang{;;98}, which translate to physical axes of 0.61 Mpc and 0.46, respectively, at the redshift of the cluster. We chose as the background region a circular region on the same chip (Chip 3) where no emission was evident. 
We used the $\texttt{dmextract}$ functionality of CIAO to estimate source counts and errors. We used the Sherpa package \citep{2001SPIE.4477...76F} to fit an \texttt{apec} component (thermal bremsstrahlung emission) $\times$ \texttt{phabs} component (galactic absorption) to the spectrum. To perform the spectral fitting, we fixed the redshift of the cluster at 0.31 and the metallicity at 0.30 solar. We adopt the elemental abundance table of \cite{1989GeCoA..53..197A} throughout our spectral analysis. This model gave an estimate for $T_X$ equal to $8.2 \pm1.2$ keV. We used the Sherpa package to convert the photon counts to a flux estimate (5.09 $\pm$ 0.25 $\times$ $10^{-12}$ erg  $\times$ $s^{-1}$). We used \cite{2006PASP..118.1711W} to convert the redshift to a luminosity distance to obtain a value for $L_X$ of ($7.29 \pm0.19 \times 10^{44}$ erg  $\times$ $s^{-1}$) in the range 0.5-10.0 keV.  We show the best fit curve for the X-ray $\textit{Chandra}$ spectrum of the Champagne Cluster in Figure \ref{fig:spectrum}.

\subsection{$\textit{ClusterPyXT}$}

Temperature maps provide crucial insights into the physical processes shaping galaxy clusters, such as shocks, turbulence, and cooling flows. In this study, we followed the procedure outlined by \citet{Alden_2019} to generate high-resolution temperature map for Abell 2355 using $\textit{ClusterPyXT}$, a pipeline that uses adaptive circular binning (ACB) and is optimized for $\textit{Chandra}$ observations. $\textit{ClusterPyXT}$ takes observation ID(s), redshift, and hydrogen column density as inputs (we used the same values of the previous section), in addition to a desired signal-to-noise ratio (SNR). We set the SNR at 50. While the pipeline requires minimal user interaction, each step was carefully inspected to ensure accuracy. Key tasks included lightcurve extraction, flare removal, and the exclusion of point sources. $\textit{CIAO-4.16}$, utilizing $\textit{specextract}$, was employed to generate the requisite response files. Subsequently, $\mathcal{O}(10^4)$ spectral fits were performed to construct the ACB temperature map that we compare to hydrodynamic simulations in $\S$\ref{sec:simul}.

\section{\textit{XMM-Newton} data reductions and analysis}

The cluster was observed by XMM-\textit{Newton} on 2010-06-05 (ObsID 0650382501, P.I. Allen). The exposure times were 6728, 6741, and 4855 s for the M1, M2, and PN detectors, respectively. 
We used the XMM-\textit{Newton} Science Analysis System (\texttt{SAS} version 19.0.0.) to extract a global temperature and X-ray luminosity for the cluster. As the $\textit{Chandra}$ observation has a longer exposure time and better resolution, we use the XMM-\textit{Newton} data as a consistency check on our results from $\textit{Chandra}$.

We filtered out time intervals contaminated by flare events by imposing a limit of 0.3 (0.6) counts s$^{-1}$ in the 10-12 keV band in the MOS (PN) detectors. Point sources were detected with the \texttt{SAS} task \texttt{edetect\_chain} and masked out. We chose an identical aperture to the one we used for the analysis of the $\textit{Chandra}$ data, and we selected single-to-quadruple events from MOS and single-to-double events from the PN. We subtracted the cosmic and non-cosmic X-ray backgrounds using the double subtraction method described in \citet{refId1} and blank-sky event lists from stacked, source-removed EPIC observations \citep{refId2} available in the XMM-\textit{Newton} blank sky file repository\footnote[1]{https://xmm-tools.cosmos.esa.int/external/xmm\_calibration/\\
background/bs\_repository/blanksky\_all.html}. We defined a background region as a circular aperture, with a $\ang{;;70}$ radius, to the SE of the cluster. All event lists were corrected for vignetting using the \texttt{evigweight} task.

After obtaining spectra for the three EPIC detectors, we performed a simultaneous fit using \textsc{XSPEC} \citep{Arnaud21996}. We used an \texttt{apec} model for the ICM emission and a \texttt{phabs} model for the galactic absorption. We fixed the redshift at 0.308 and the HI column density at $1.08 \times 10^{20} $ cm$^{-2}$. Our results indicate an unabsorbed luminosity in the $0.5$--$10.0$ keV energy range of $L_X=6.65_{-0.34}^{+0.35} \times 10^{44}$ erg  $\times$ $s^{-1}$ and a global temperature $T_X=5.61_{-0.71}^{+0.93}$ keV, where the error bars correspond to the 90\% confidence interval. Our measurements are consistent with our results from $\textit{Chandra}$ described above to within 2$\sigma$.

Based on the richness-$T_X$ scaling relationship presented by \citet{Rykoff_Rozo}, we derive a value of $\sim$ 5 keV for $T_X$ for the Champagne Cluster, with a scatter of up to 40$\%$, based on its richness. The values we derive using both $\textit{Chandra}$ and XMM-$\textit{Newton}$ are consistent with this estimate.

\section{Spectroscopic data analysis}

\begin{figure*}
  \centering
  \includegraphics[width=180mm]{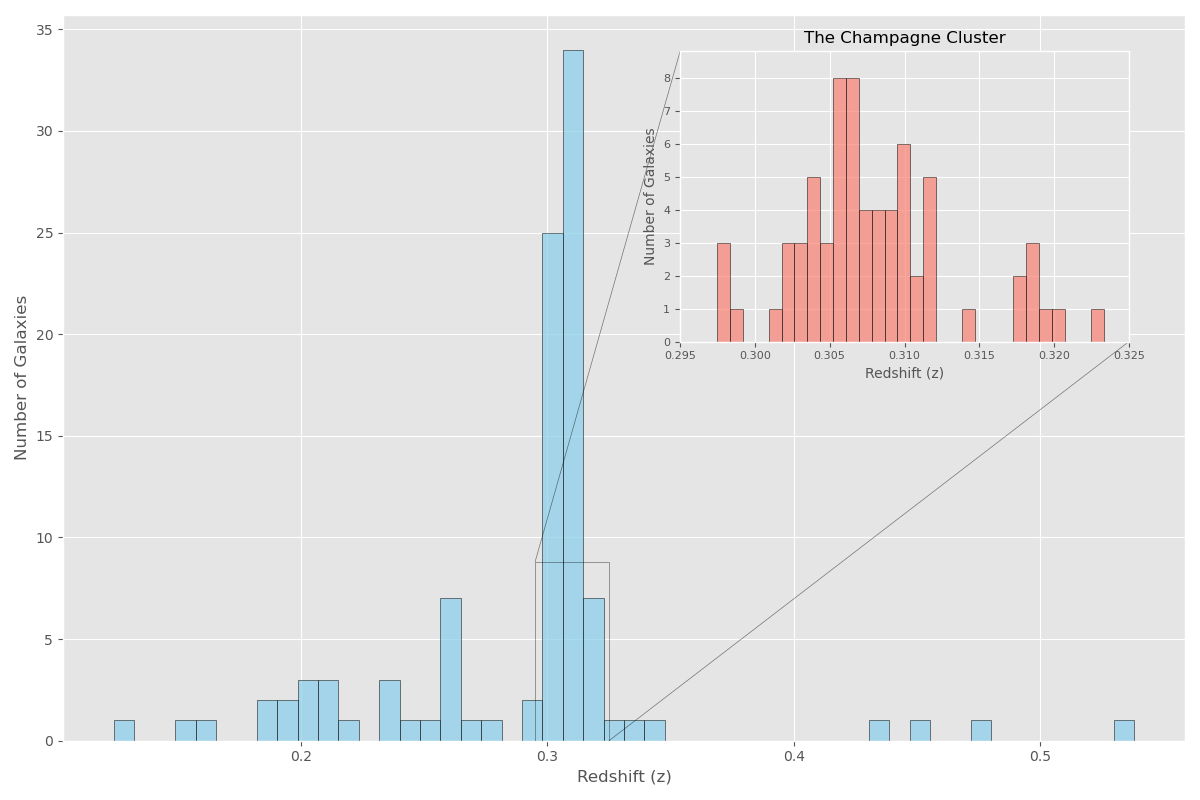}
  \caption{Members of the spectroscopic redshift catalog, with inset showing redshift distribution of potential cluster member galaxies for the Champagne Cluster.}
  \label{fig:redshifts}
\end{figure*}

\normalsize\textit{Data acquisition:} We observed the Champagne Cluster with the DEIMOS multi-object spectrograph \citep{FaberDEIMOS} at the W. M. Keck Observatory on 2022 July 1 (UT). Because the DEIMOS field of view is approximately $16^{\prime} \times 4^{\prime}$, it is ideally suited for observing merging clusters, particularly when the mask's long axis is placed parallel to the merger axis. We used two slitmasks with a total of 150 (80 and 70, respectively) $1''$ wide slits. We selected potential member galaxies based on Pan-STARRS photometric redshifts \citep{2021MNRAS.500.1633B}. Each galaxy in the Pan-Starrs photometric redshift catalog has a redshift uncertainty of $\sigma_{\textrm{PS}}$. The likelihood of a given galaxy to be at a cluster at redshift $z_{\textrm{cl}}$ is given by

\bigskip
\[
\ell \propto \frac{1}{\sigma_{\textrm{PS}}} \exp \left( \frac{(z_{\textrm{PS}} - z_{\textrm{cl}})^2}{2 \sigma_{\textrm{PS}}^2} \right)
\]
\bigskip

\noindent The typical photometric redshift uncertainty of 0.16 enables some focus on the target cluster while also probing foreground and background structures. Additional priority was given to brighter targets.

We used the 1200 line mm$^{\textrm{-1}}$ grating, which results in a pixel scale of 0.33 $\textrm{\r{A}}$ pixel$^{-1}$ and a resolution of $\sim$ 1$\textrm{\r{A}}$ (50 km/s).
The grating was tilted to observe the wavelength range $\approx$ 4200-6900 $\textrm{\r{A}}$ (the precise range depends on the slit position), which at z=0.308 includes spectral features from the [OII] 3727 $\textrm{\r{A}}$ doublet to the magnesium line at 5177 $\textrm{\r{A}}$. The total exposure time for each mask was 27 minutes divided into three exposures. 

$\normalsize\textit{Data reduction and redshift extraction}$: We used PypeIt \citep{2020zndo...3743493P} to calibrate the data and reduce it to a series of 1D spectra. We developed a Python software (Pyze) to extract redshifts from 1D spectra. Pyze adopts the approach of the DEEP2 survey \citep{2013ApJS..208....5N}, which removes a slowly varying empirical continuum model before performing real-space cross-correlation with templates. 

This software is described in more detail in \citet{Wittman_2023}. Since that paper, we have reduced user-to-user redshift variations to the $10^{-5}$ level by making the redshift refinement more independent of the initial search parameters. The uncertainty is now calculated purely from the curvature of the $\chi^{2}$ surface, typically yielding uncertainties of $\leq$ $10^{-4}$ in redshift or $\leq$ 23 km/s in the frame of the cluster.  This does not capture uncertainties in wavelength calibration. These are difficult to quantify, but repeat observations suggest they are at the $10^{-5}$ level. Table ~\ref{table:tab-z} lists the redshifts we obtained through the Keck/DEIMOS run.
\startlongtable
\begin{deluxetable}{llll}
  \tablecaption{Observed redshifts}  \label{table:tab-z}
  \tablecolumns{4}
  \tablehead{\colhead{R.A. (deg)} & \colhead{Decl. (deg)}  &\colhead{z} & \colhead{Uncertainty in z}}
  \startdata
196.327725  &26.461019  &0.305962  &0.000234\\
196.336867  &26.436628  &0.303811  &0.000033\\
196.340367  &26.468936  &0.437909  &0.000101\\
196.347330  &26.435890  &0.256980  &0.000024\\
196.354983  &26.437925  &0.305562  &0.000096\\
196.355500  &26.483536  &0.308447  &0.000165\\
196.357233  &26.511017  &0.304311  &0.000014\\
196.369908  &26.466914  &0.255032  &0.000029\\
196.373521  &26.471492  &0.213884  &0.001472\\
196.377208  &26.503956  &0.306413  &0.000048\\
196.377800  &26.468908  &0.258414  &0.000034\\
196.380112  &26.469231  &0.220285  &0.000012\\
196.385780  &26.495860  &0.303828  &0.000037\\
196.404483  &26.496414  &0.303311  &0.000122\\
196.410330  &26.494960  &0.184248  &0.000025\\
196.413475  &26.468006  &0.303721  &0.000061\\
196.419221  &26.511228  &0.308841  &0.000028\\
196.424454  &26.460528  &0.309298  &0.000072\\
196.424792  &26.513903  &0.307623  &0.000059\\
196.439867  &26.509047  &0.303974  &0.000038\\
196.441833  &26.514008  &0.309431  &0.000042\\
196.457700  &26.510725  &0.318771  &0.000082\\
196.462067  &26.492536  &0.306313  &0.000031\\
196.475380  &26.502540  &0.297723  &0.000118\\
196.478480  &26.501668  &0.309465  &0.000085\\
196.480583  &26.533928  &0.256971  &0.000016\\
196.482008  &26.524608  &0.307964  &0.000247\\
196.482921  &26.508169  &0.317520  &0.000050\\
196.484062  &26.512369  &0.323307  &0.000078\\
196.485979  &26.514753  &0.305912  &0.000029\\
196.499188  &26.565442  &0.244575  &0.000020\\
196.502396  &26.514214  &0.311783  &0.000049\\
196.505046  &26.515486  &0.309932  &0.000104\\
196.509163  &26.569164  &0.207280  &0.000057\\
196.514287  &26.522178  &0.197854  &0.000039\\
196.515092  &26.525225  &0.199591  &0.000008\\
196.516858  &26.541944  &0.298207  &0.000049\\
196.530100  &26.590400  &0.298491  &0.000061\\
196.530583  &26.594114  &0.337267  &0.000046\\
196.534490  &26.548370  &0.301142  &0.000098\\
196.536404  &26.515406  &0.311966  &0.000067\\
196.542904  &26.554075  &0.304628  &0.000121\\
196.545617  &26.582072  &0.258731  &0.000017\\
196.556692  &26.544931  &0.310048  &0.000041\\
196.561896  &26.572611  &0.258257  &0.000046\\
196.568742  &26.576467  &0.305429  &0.000115\\
196.598842  &26.555558  &0.237650  &0.000034\\
196.406071  &26.507350  &0.306413  &0.000096\\
196.407129  &26.478233  &0.307814  &0.000078\\
196.412379  &26.482011  &0.309665  &0.000040\\
196.413208  &26.522172  &0.305237  &0.000040\\
196.418250  &26.482703  &0.307438  &0.000040\\
196.419308  &26.490708  &0.303361  &0.000138\\
196.420488  &26.498347  &0.305826  &0.000020\\
196.422767  &26.470219  &0.320555  &0.000049\\
196.426300  &26.484389  &0.306604  &0.000049\\
196.430721  &26.501328  &0.311833  &0.000045\\
196.434650  &26.536753  &0.310499  &0.000035\\
196.442446  &26.515369  &0.306663  &0.000081\\
196.447192  &26.487183  &0.276126  &0.000015\\
196.453525  &26.481178  &0.302410  &0.000048\\
196.457129  &26.487908  &0.124041  &0.000018\\
196.460138  &26.475936  &0.310589  &0.000016\\
196.461229  &26.537186  &0.307213  &0.000023\\
196.461463  &26.509986  &0.306062  &0.000079\\
196.466996  &26.535747  &0.318054  &0.000039\\
196.467113  &26.531742  &0.197890  &0.000004\\
196.468504  &26.512839  &0.304628  &0.000077\\
196.473296  &26.530936  &0.319471  &0.000022\\
196.478412  &26.501717  &0.310132  &0.000075\\
196.489467  &26.509808  &0.306763  &0.000046\\
196.495425  &26.513433  &0.308097  &0.000027\\
196.496883  &26.504150  &0.340836  &0.000039\\
196.498125  &26.515661  &0.311543  &0.000079\\
196.500612  &26.517731  &0.310182  &0.000096\\
196.502837  &26.520478  &0.306873  &0.000097\\
196.506196  &26.529386  &0.236583  &0.000228\\
196.511362  &26.516389  &0.304895  &0.000029\\
196.516858  &26.521828  &0.318504  &0.000141\\
196.519875  &26.551706  &0.207438  &0.000141\\
196.521813  &26.519706  &0.311466  &0.000120\\
196.522946  &26.529539  &0.314285  &0.000041\\
196.524037  &26.531869  &0.309698  &0.000032\\
196.533392  &26.539300  &0.538167  &0.000065\\
196.535112  &26.519197  &0.297473  &0.000056\\
196.549442  &26.566467  &0.155595  &0.000010\\
196.549825  &26.574411  &0.305962  &0.000045\\
196.557217  &26.537475  &0.306629  &0.000106\\
196.558154  &26.525625  &0.301910  &0.000059\\
196.560404  &26.597644  &0.301960  &0.000068\\
196.563408  &26.582533  &0.236666  &0.000073\\
196.569979  &26.576467  &0.451659  &0.000122\\
196.574483  &26.610456  &0.257530  &0.000079\\
196.575971  &26.555606  &0.303461  &0.000014\\
196.606979  &26.547783  &0.203394  &0.000155\\
196.609442  &26.585897  &0.205228  &0.000084\\
196.612413  &26.556906  &0.473916  &0.000033\\
196.622983  &26.623392  &0.185849  &0.000025\\
196.632420  &26.560800  &0.307513  &0.000020\\
196.651542  &26.574550  &0.158923  &0.000377\\
196.669200  &26.584300  &0.267854  &0.000034\\
196.671800  &26.602144  &0.318411  &0.000014\\

    \enddata
\end{deluxetable}

\normalsize\textit{Archival redshifts acquisition:}
\label{subsec:archival}
We found 22 spectroscopic redshifts in NED for galaxies within $10^{\prime}$ from the position of the most likely BCG candidate of the Champagne Cluster. We obtained two additional unique redshifts from the Early Data Release of the Dark Energy Spectroscopic Instrument \citep{DESI}. This gave us a total of 24 unique archival redshifts. After contrasting the list of 24 unique archival redshift against our Keck/DEIMOS catalog, we found one duplicate. The redshift and velocity difference between the duplicate archival redshift and the corresponding entry in our Keck/DEIMOS catalog is 3.95$\times$$10^{-5}$ and 12.8 km/s, respectively. After removing the duplicate redshift, we added the remaining 23 redshifts to the {\bfseries 102} redshifts from our Keck/DEIMOS spectroscopic survey to form a final catalog of {\bfseries 125} unique redshifts, which we use in the analysis we undertake in $\S$\ref{sec:dynamics}.

\section{Subclustering and kinematics}
\label{sec:dynamics}

\begin{figure*}
  \centering
  \includegraphics[width=180mm]{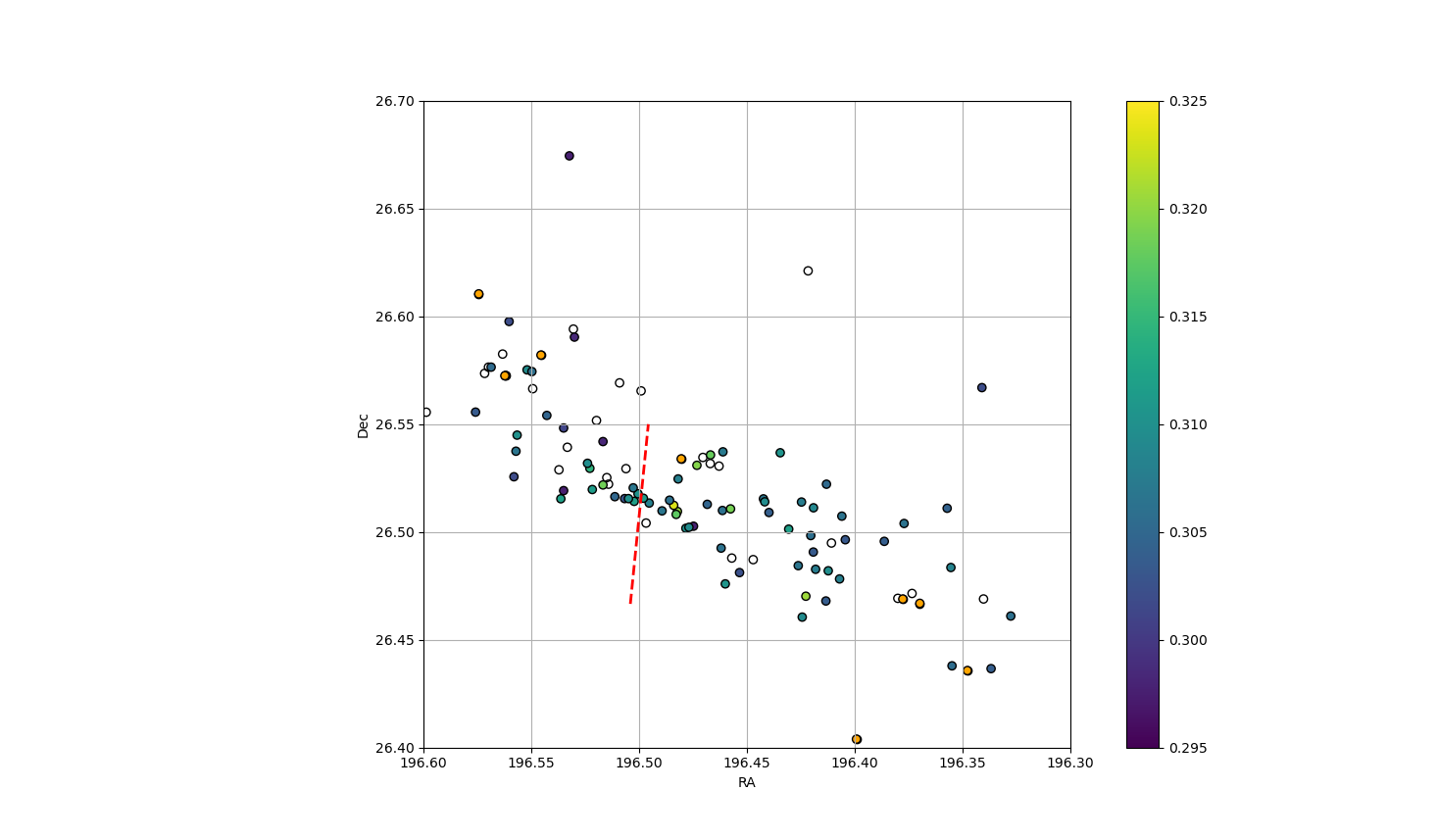}
  \caption{Redshift distribution for the members of the spectroscopic catalog:  The Champagne Cluster (color bar), galaxies in the redshift range 0.250 $\leq$ z $\leq$ 0.270 (orange), and the remaining galaxies (open circles). The dashed line separates the two subclusters, Champagne-NW and Champagne-SE.}
  \label{fig:foreground}
\end{figure*}

We present the members of the spectroscopic redshift catalog in Figure~\ref{fig:redshifts}. The histogram indicates the presence of a foreground peak of galaxies at $z$ = 0.260. This group has a mean redshift of 0.2575 $\pm$ 0.0004 and a rest-frame velocity dispersion of 361 $\pm$ 96 km/s. Figure \ref{fig:foreground} shows the locations of all the galaxies in the spectroscopic catalog. We do not see a spatial concentration of the points at z=0.26, so we conclude it is not a foreground cluster that would complicate our analysis.

To analyze the dynamics of the Champagne Cluster, we restricted the redshift range to a small window (0.295$\leq$ z $\leq$ 0.325) centered on the redshifts of the cluster's top BCG candidate (a total of 75 galaxies). We used the biweight estimator \citep{1990AJ....100...32B} to calculate the systemic redshift for the Champagne Cluster (0.3069 $\pm$ 0.0005) and the velocity dispersion in the cluster rest-frame (1204 $\pm$ 155 km/s). We conducted the Anderson-Darling test on the redshift population of the Champagne Cluster to evaluate its consistency with a Gaussian distribution. The resulting p-value of 0.15 indicates that the redshift population of the Champagne Cluster does not deviate significantly from Gaussianity.

\begin{figure}
  \centering
  \includegraphics[width=\columnwidth]{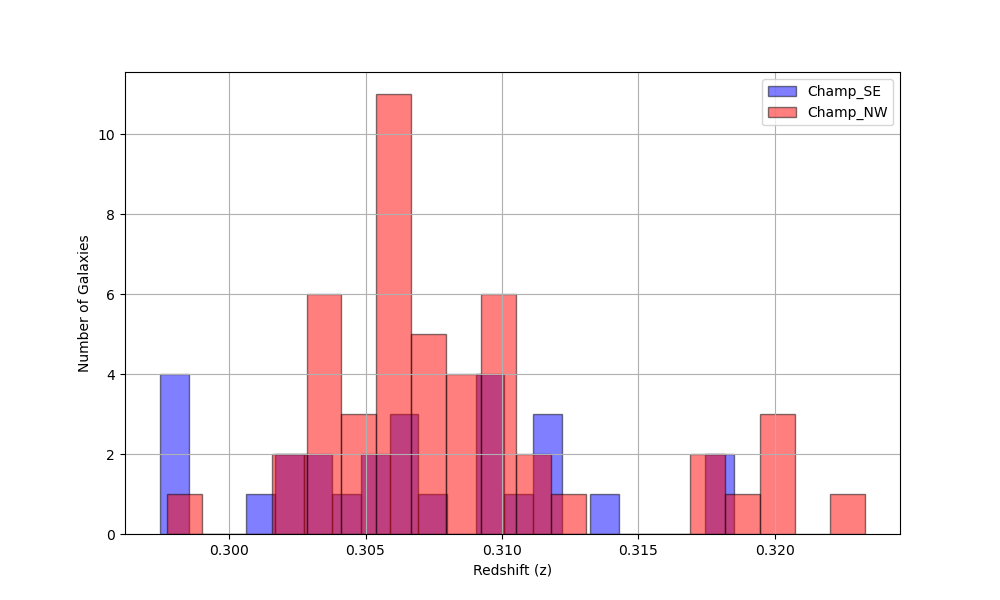}
  \caption{The two subclusters of the Champagne Cluster: Champagne-SE and Champagne-NW.}
  \label{fig:substructure}
\end{figure}

We defined Champagne-SE and Champagne-NW as the two clusterings of galaxies appearing, respectively, to the southeast and northwest of the X-ray surface brightness peak as indicated by the dashed-red line in Figure~\ref{fig: Champagne}. Champagne-SE (Champagne-NW) has 27 (48) spectroscopic galaxy redshifts. We used the biweight method to obtain a systemic redshift for Champagne-SE (Champagne-NW) of $0.3065 \pm 0.0012$ ($0.3069 \pm 0.0005$) and a velocity dispersion of 1357$\pm$171 km/s (1015$\pm$242 km/s). Figure~\ref{fig:substructure} illustrates the redshift population of each subcluster. We performed the Anderson-Darling test on the redshift population of each subcluster. The p-value for Champagne-SE is 0.73, indicating that the redshift population fits a single Gaussian distribution well. Champagne-NW has a p-value of 5.15 $\times$ $10^{-5}$, suggesting that it significantly deviates from Gaussianity. Although the system is a binary merger based on its X-ray morphology, large BCG separation, and optical substructure, the velocity distribution of cluster members is well described by a single Gaussian model. This likely reflects the orientation of the merger axis near the plane of the sky, which minimizes the line-of-sight velocity difference between the subclusters. Such behavior is commonly observed in dissociative merging systems where the merger geometry suppresses velocity bimodality \citep{2006ApJ...648L.109C}.

We used the average redshift of each subcluster as a proxy for the redshift of the subcluster as a whole to refine the value of the velocity of one cluster in the frame of the other, $\Delta v_{\text{los}}$. We get a value of 411 $\pm$ 180 km/s. This value is larger than the value we obtained in $\S$\ref{sec:overview}, using the BCGs alone. This value is consistent with the value derived from the BCGs alone at the 2.3$\sigma$ level, albeit with large uncertainties due to the large velocity dispersion of the member galaxies. It is possible that the BCGs are better tracers of the central part of the potential, but it is difficult to estimate the uncertainty in the relative velocity of the BCGs alone. Therefore, in the modeling that follows we use the result from all member galaxies to avoid overconfidence in the modeling.

\label{section:Manual}

\section{Numeric Simulations}
\subsection{Simulated analogs}
\label{sec:analogs}
We searched for analogs to the Champagne Cluster in the Big Multidark Planck Simulation (BigMDPL) \citep{Klypin} using the method of \citet{Wittman_2018} and \citet{Wittman_2019} with one improvement as follows.  \citet{Wittman_2019} noted that the velocities listed in the halo catalog are often underestimated when the halos are largely overlapping, resulting in the pericenter speed being biased low. Therefore, they estimate the pericenter speed as the maximum relative halo speed over all snapshots near pericenter. However, we found that this method tends to underestimate the true pericenter speed, likely due to limited time resolution in the simulation outputs. To address this, we now compute the pericenter speed by directly measuring the change in relative position over time between the two halos across adjacent snapshots. This interpolation yields a more accurate estimate and increases the inferred pericenter speed by approximately 200–400 km/s compared to the original method.

We used the line-of-sight relative velocity ($\Delta v_{21}$) that we calculated in $\S$\ref{section:Manual}, and we used the separation between the top two BCG candidates from $\S$\ref{sec:overview} as a proxy for the separation between the two subclusters. BCGs usually reside near the bottom of the gravitational well, but it is difficult to quantify how near so we assume a generous 100 kpc uncertainty.

In Table \ref{tab:dynamic}, we list the {\bfseries 95$\%$} confidence interval for the time since pericenter (TSP), pericenter speed $v_{\text{max}}$, viewing angle $\theta$ (defined as the angle the subcluster separation vector makes with the line of sight), and $\varphi$ (defined as the angle the current separation vector makes with the velocity vector) for outbound and returning systems. Qualitatively, the relatively low line-of-sight velocity, $\Delta v_{21}$, suggests two possibilities: a merger happening near the plane of the sky and/or a merger at turnaround. The combined values of $v_{\text{max}}$ and $\theta$ (Table \ref{tab:dynamic}) seem to prefer a merger near the plane of the sky.  

\subsection{Hydrodynamic Simulations} 
\label{sec:simul}
To constrain the scenario further, we used the hydrodynamic, binary merger simulations from the Galaxy Cluster Merger Catalog (http://gcmc.hub.yt/) \citep{ZuHone}. We used different scenarios with different combinations of mass ratios (1:1, 1:3, and 1:10), impact parameters (0 kpc, 500 kpc, and 1,000 kpc), and viewing angles (x, y, and z), where x is the direction of motion (initial velocity vector), y is the direction perpendicular to x in the plane of motion, and z is the direction perpendicular to the plane of motion. We compared the location of the peak of X-ray emissivity between the BCGs, temperature map, and the locations of the BCGs (we assumed the BCGs are located at the peaks of the total mass density) to our data. The scenarios with mass ratios of 1:1 and 1:3 produced gas temperatures hotter than what we observed. The scenario with mass ratio of 1:10 and impact parameter of 1 Mpc produced a temperature that was cooler than what was observed. There are two remaining possibilities: a 0 impact parameter with a mass ratio of 1:10 and a 500 kpc impact parameter with a mass ratio of 1:10.

\begin{figure*}
  \centering
  \includegraphics[width=170mm]{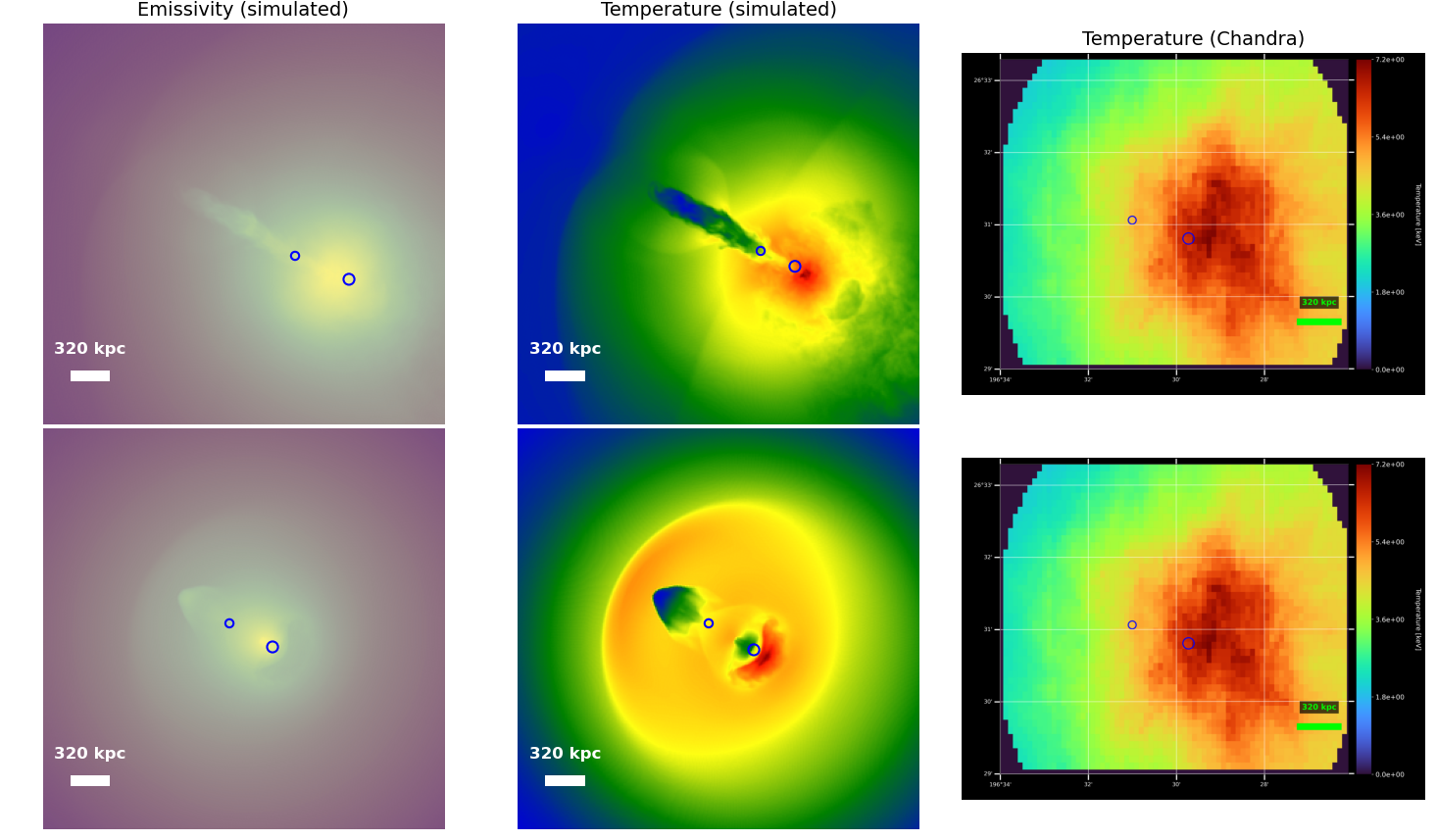}
  \caption{Comparison of hydrodynamic simulations Galaxy Cluster Merger Catalog (http://gcmc.hub.yt/) \citep{ZuHone} and a $\textit{Chandra}$ $\textit{ClusterPyXT}$ temperature map. Scenario I (top panel): an impact parameter of 0 kpc, a mass ratio of 1:10, and a TSP of 2.20 Gyr. Scenario II (bottom panel): an impact parameter of 500 kpc, a mass ratio of 1:10, and a TSP of 0.4 Gyr. The hot gas is to the right of the dominant BCG (large blue circle) in both simulations and in the $\textit{Chandra}$ temperature map. The temperature ranges from 0 to 7.2 keV in the $\textit{Chandra}$ temperature map as well as in the simulations. The peak of the X-ray emissivity is located on the BCG separation vector, between the two BCGs (colored circles), closer to the dominant BCG, as in Figure~\ref{fig: Champagne}}.
  \label{fig:simulations}
\end{figure*}

These possibilities produced two scenarios that corresponded well to the data: 1) a returning system viewed along the z-direction with an impact parameter of 0 and a TSP of 2.2 Gyr (scenario I), and 2) an outbound system viewed along the x-direction with an impact parameter of 500 kpc and a TSP of 0.4 Gyr (scenario II). Scenario I predicts a $\Delta$$v_{los}$ of 0 km/s, and scenario II predicts a $\Delta$$v_{los}$ of 743 km/s. While both scenarios are broadly consistent with the observed temperature and X-ray surface brightness density maps, and the $\Delta$$v_{los}$ we derived in $\S$\ref{sec:dynamics}, scenario I gives a better agreement with the observed data if we use the $\Delta$$v_{los}$ of the BCGs (6 $\pm$ 20 km/s) that we derived in $\S$\ref{sec:overview} as a proxy for the $\Delta$$v_{los}$ of the two clusters. The viability of scenario II depends on choosing a particular line of sight (the x-direction), while the viability of scenario I is compatible with two lines of sight (y and z). Both scenarios indicate the more massive subcluster is closer to the region with the largest temperature, which is consistent with the data (the region with the hottest temperature is closed to the cluster with the most dominant BCG, Champagne-SE). We present the simulations and temperature map side by side in Figure~\ref{fig:simulations}. We observe that the TSP is lower for the analogs than the hydrodynamic simulations, both for outbound and inbound systems. A better estimation of the mass of the Champagne Cluster through a weak-lensing analysis can shed more light on the merger timescale.

\begin{table}
  \centering  
  \caption{Dynamical parameters from analogs}
  \label{tab:dynamic}
  \begin{tabular}{cccc}
    \hline
    TSP (Myr) & $v_{\text{max}}$ (km/s) & $\theta$ (deg) & $\varphi$ (deg) \\
    \hline
    \multicolumn{4}{c}{Outbound only} \\ 
    65-231 & 1568-2112 & 62-90 &2-30 \\ 
    \hline
    \multicolumn{4}{c}{Returning only} \\
    1253-2010 & 2089-2607 & 58-90 & 156-174 \\ 
    \hline
  \end{tabular}
\end{table}

 \section{Conclusions}

We have presented a new bimodal galaxy cluster merger, the Champagne Cluster, with a number of remarkable features. The offset between the X-ray peak and the BCGs of the subclusters suggests that the collision trajectory was near head-on. The low line of sight velocity difference between the two subclusters involved in the merger (Champagne-SE and Champagne-NW), in addition to simulated analogs {\bfseries and hydrodynamic simulations}, indicate the merger is happening mainly in the plane of the sky. These features make the Champagne Cluster a promising system for follow-up as it can potentially shed some more light on the reaction of dark matter to a high speed collision. High-resolution weak lensing and X-ray maps could constrain the dynamics further by providing accurate estimates for the masses of the subclusters, as well as offsets between the mass centroids and the X-ray emission peak and the presence of shocks and temperature or density discontinuities. 

\section{acknowledgements}

This work was supported by NSF grant 2308383.

This research has made use of data obtained from the $\textit{Chandra}$ Data Archive and the $\textit{Chandra}$ Source Catalog, and of \textit{Aladin sky atlas} developed at CDS, Strasbourg Observatory, France \citep{2000A&AS..143...33B}. It has also made use of software provided by the $\textit{Chandra}$ X-ray Center (CXC) in the application packages CIAO-4.14 \citep{2006SPIE.6270E..1VF} and Sherpa \citep{2001SPIE.4477...76F}.

This research has also made use of data data obtained from the Legacy Surveys. The Legacy Surveys consist of three individual and complementary projects: the Dark Energy Camera Legacy Survey (DECaLS; Proposal ID :2014B-0404; PIs: David Schlegel and Arjun Dey), the Beijing-Arizona Sky Survey (BASS; NOAO Prop. ID:015A-0801; PIs: Zhou Xu and Xiaohui Fan), and the Mayall z-band Legacy Survey (MzLS; Prop. ID :2016A-0453; PI: Arjun Dey). DECaLS, BASS and MzLS together include data obtained, respectively, at the Blanco telescope, Cerro Tololo Inter-American Observatory, NSF’s NOIRLab; the Bok telescope, Steward Observatory, University of Arizona; and the Mayall telescope, Kitt Peak National Observatory, NOIRLab. Pipeline processing and analyses of the data were supported by NOIRLab and the Lawrence Berkeley National Laboratory (LBNL). The Legacy Surveys project is honored to be permitted to conduct astronomical research on Iolkam Du’ag (Kitt Peak), a mountain with particular significance to the Tohono O’odham Nation.

NOIRLab is operated by the Association of Universities for Research in Astronomy (AURA) under a cooperative agreement with the National Science Foundation. LBNL is managed by the Regents of the University of California under contract to the U.S. Department of Energy.

This project used data obtained with the Dark Energy Camera (DECam), which was constructed by the Dark Energy Survey (DES) collaboration. Funding for the DES Projects has been provided by the U.S. Department of Energy, the U.S. National Science Foundation, the Ministry of Science and Education of Spain, the Science and Technology Facilities Council of the United Kingdom, the Higher Education Funding Council for England, the National Center for Supercomputing Applications at the University of Illinois at Urbana-Champaign, the Kavli Institute of Cosmological Physics at the University of Chicago, Center for Cosmology and Astro-Particle Physics at the Ohio State University, the Mitchell Institute for Fundamental Physics and Astronomy at Texas A\&M University, Financiadora de Estudos e Projetos, Fundacao Carlos Chagas Filho de Amparo, Financiadora de Estudos e Projetos, Fundacao Carlos Chagas Filho de Amparo a Pesquisa do Estado do Rio de Janeiro, Conselho Nacional de Desenvolvimento Cientifico e Tecnologico and the Ministerio da Ciencia, Tecnologia e Inovacao, the Deutsche Forschungsgemeinschaft and the Collaborating Institutions in the Dark Energy Survey. The Collaborating Institutions are Argonne National Laboratory, the University of California at Santa Cruz, the University of Cambridge, Centro de Investigaciones Energeticas, Medioambientales y Tecnologicas-Madrid, the University of Chicago, University College London, the DES-Brazil Consortium, the University of Edinburgh, the Eidgenossische Technische Hochschule (ETH) Zurich, Fermi National Accelerator Laboratory, the University of Illinois at Urbana-Champaign, the Institut de Ciencies de l’Espai (IEEC/CSIC), the Institut de Fisica d’Altes Energies, Lawrence Berkeley National Laboratory, the Ludwig Maximilians Universitat Munchen and the associated Excellence Cluster Universe, the University of Michigan, NSF’s NOIRLab, the University of Nottingham, the Ohio State University, the University of Pennsylvania, the University of Portsmouth, SLAC National Accelerator Laboratory, Stanford University, the University of Sussex, and Texas A\&M University.

BASS is a key project of the Telescope Access Program (TAP), which has been funded by the National Astronomical Observatories of China, the Chinese Academy of Sciences (the Strategic Priority Research Program ``The Emergence of Cosmological Structures” Grant $\#$ XDB09000000), and the Special Fund for Astronomy from the Ministry of Finance. The BASS is also supported by the External Cooperation Program of Chinese Academy of Sciences (Grant $\#$ 114A11KYSB20160057), and Chinese National Natural Science Foundation (Grant $\#$ 12120101003, $\#$ 11433005).

\normalsize\textit{Facilities:} Keck:II (Deimos), Chandra, XMM.

This work made use of data from the Galaxy Cluster Merger Catalog (http://gcmc.hub.yt/)



\end{document}